\documentclass[double]{article}
\usepackage{times}
\usepackage{algorithm}
\usepackage{algorithmic}
\usepackage{wrapfig}
\usepackage{verbatim}
\usepackage{amsmath}
\usepackage{graphicx}
\usepackage{subcaption}
\usepackage[active]{srcltx}
\usepackage{color}
\usepackage{lineno}
\usepackage{dirtytalk}
\usepackage{amsthm}
\usepackage{authblk}
\usepackage{listings}
\usepackage{tikz}

\usepackage{epsfig,graphicx,amsfonts}
\usepackage{amsmath,amsthm,amssymb}
\usepackage{xcolor}

\usepackage{adjustbox}
\usepackage{multirow}
\usepackage{setspace}
\usepackage{hyperref}
\usepackage{comment}

\long\def\comment #1\commentend{}

\begin{document}

\title{\Large Temporal Graphs Anomaly Emergence Detection: Benchmarking For Social Media Interactions}
\author{Teddy Lazebnik$^{1}$ and Or Iny$^{2}$\\
\(^1\) Department of Cancer Biology, Cancer Institute, University College London, London, UK\\
\(^2\) Department of Economy, The Academic College of Tel Aviv–Yaffo, Tel Aviv, Israel\\
\(*\) Corresponding author: lazebnik.teddy@gmail.com

}

\date{}

\maketitle 

\begin{abstract}
Temporal graphs have become an essential tool for analyzing complex dynamic systems with multiple agents. Detecting anomalies in temporal graphs is crucial for various applications, including identifying emerging trends, monitoring network security, understanding social dynamics, tracking disease outbreaks, and understanding financial dynamics. In this paper, we present a comprehensive benchmarking study that compares 12 data-driven methods for anomaly detection in temporal graphs. We conduct experiments on two temporal graphs extracted from Twitter and Facebook, aiming to identify anomalies in group interactions. Surprisingly, our study reveals an unclear pattern regarding the best method for such tasks, highlighting the complexity and challenges involved in anomaly emergence detection in large and dynamic systems. The results underscore the need for further research and innovative approaches to effectively detect emerging anomalies in dynamic systems represented as temporal graphs.
\end{abstract}

\maketitle \thispagestyle{empty}

\pagestyle{myheadings} \markboth{Draft:  \today}{Draft:  \today}
\setcounter{page}{1}

\section{Introduction}
\label{sec:introduction}
The analysis of complex dynamic systems with multiple agents has gained significant attention in various fields, such as social networks \cite{intro_social}, biological systems \cite{intro_bio}, and transportation networks \cite{intro_cars}. Recently, temporal graphs have gained much attention as a fundamental framework for capturing the dynamic nature of these systems, enabling the study of evolving relationships and interactions over time \cite{tg_tag_1,tg_tag_2,tg_tag_3,tg_tag_4}. Representing systems as temporal graphs is considered straightforward in most cases which makes it a robust and appealing data structure to use \cite{tg_math_1}. 

Anomalies in temporal graphs can manifest as unexpected shifts in network behavior, sudden changes in interaction patterns, or the emergence of unusual group dynamics \cite{tg_anomaly_1,tg_anomaly_2}. These anomalies often provide valuable insights into significant events, emerging phenomena, or potentially malicious activities within the underlying system. Detecting emerging anomalies in such temporal graphs has become a critical task with wide-ranging applications, including identifying credit frauds \cite{tg_ano_exp_1}, identifying social trends \cite{tg_ano_exp_2}, and understanding cell-level biological processes \cite{tg_ano_exp_3}. Consequently, developing effective methods for anomaly emergence detection in temporal graphs allows temporally-close-proximity or even immediate reaction to shifts in the dynamics.

Several approaches have been proposed to tackle the challenge of anomaly emergence detection in general \cite{ano_task_1,ano_task_2}, and in temporal graphs, in particular \cite{ano_task_3,ano_task_4}. These approaches span statistical methods, machine learning algorithms, and graph-based techniques, each leveraging different assumptions and models to capture the unique characteristics of temporal graph data \cite{ano_task_5,ano_task_6}. However, due to the complexity and inherent uncertainty associated with detecting anomalies in dynamic systems, identifying the most suitable method for a specific application remains mostly unclear.

In this paper, we present a comprehensive benchmarking study that focuses on the task of anomaly emergence detection in temporal graphs, with a specific emphasis on social media interactions. Social media platforms, such as Twitter and Facebook, provide rich sources of temporal graph data, capturing the dynamic interactions among individuals, groups, and communities that can shed light on social and economic trends in real-time. Detecting anomalies in group interactions within these platforms holds immense value in understanding influential events, collective behaviors, and the spread of information. In particular, we evaluated 12 state-of-the-art methods that represent a diverse range of approaches and techniques employed in the field. By conducting experiments on two temporal graphs obtained from Twitter and Facebook, we seek to investigate the performance of these methods in identifying anomalies in group interactions within the context of social media.

Our findings present an unexpected outcome: an unclear pattern emerges regarding the best-performing method for anomaly emergence detection in social media interactions. This outcome underscores the need for further research and the development of novel techniques tailored to the unique characteristics of social media data.

This paper is structured as follows. Section \ref{sec:related_work} provides an overview of the temporal graphs' data structure as well as the formalization of anomaly emergence detection. Next, section \ref{sec:experiment} describes the methodology and experimental setup employed in our benchmarking study.  Subsequently, section \ref{sec:results} presents the performance of each method on the Twitter and Facebook temporal graphs. Finally, section \ref{sec:discussion} analyzes our findings and suggests potential future studies.

\section{Related Work}
\label{sec:related_work}
Temporal graphs have gained significant attention in various domains as a means to capture the evolving relationships and interactions in complex dynamic systems \cite{rw_intro_1,rw_intro_2,rw_intro_3}. In this section, we provide a formalization of temporal graphs followed by the anomaly emergence detection task definition. 

Temporal (also known as dynamic, evolving, overtime-varying) graphs can be informally described as graphs that change with time. A temporal graph is a mathematical representation of a dynamic system that captures both the structural properties of a graph and the temporal aspects of interactions between entities. Formally, a temporal graph can be defined as follow. Let \(G = (V, E, T)\) be a temporal graph, where \(V \in \mathbb{N}^k\) represents the set of nodes or entities in the graph represented as finite state machines with \(k \in \mathbb{N}\) possible states, \(E \subset V \times V \times \mathbb{R}\) denotes the set of edges such that each edge \(e \in E := (u, v, t)\) represents an interaction between nodes \(u\) and \(v\) at time \(t\), and \(T \in \mathbb{N}\) is the set of discrete time points or intervals at which the interactions occur. Intuitively, one can represent a temporal graph as a set of timestamped edges, \(G = {(u, v, t) | (u, v) \in E, t \in T}\), that implicitly indicates the nodes of the graph and their interactions over time. 

Though the formal treatment of temporal graphs is still in its infancy, there is already a huge identified set of applications and research domains that motivate it and that could benefit from the development of a concrete set of results, tools, and techniques for temporal graphs \cite{tg_intro_book}. In the domain of biological systems, for instance, gene regulatory networks can be represented as temporal graphs, where nodes correspond to genes and edges capture interactions between genes at different time points, which allows the study of gene expression patterns \cite{rw_genes}. Indeed, \cite{rw_gene_exp} proposed an inference algorithm based on linear ordinary differential equations. The authors show that algorithm can infer the local network of gene–gene interactions surrounding a gene of interest from time-series gene expression profiles of synthetic genomics samples. In addition, in the transportation systems realm, nodes of a temporal graph can represent locations, and edges capture movements or interactions between locations at different time points, providing an intuitive formalization to analyze traffic flows and congestion patterns \cite{intro_cars}. For example, \cite{rw_traffic_example} propose a framework that enables extending the traditional convolutional neural network model to graph domains and learns the graph structure for traffic forecasting. Most relevant for this work, temporal graphs can capture the evolving relationships between individuals, communities, and groups over time. They enable the study of social phenomena, such as information diffusion \cite{rw_social_general_1}, opinion formation \cite{rw_social_general_2}, and community detection \cite{intro_bio}. \cite{rw_social_example} propose a dynamic graph-based framework that leverages the dynamic nature of the users’ network for detecting fake news spreaders. Using their model, the authors show that by analyzing the users’ time-evolving semantic similarities and social interactions, one can indicate misinformation spreading. 

While there are many possible queries one can perform on a temporal graph, we focus on detecting anomalies over time in close temporal proximity to when they start to emerge. Namely, the anomaly emergence detection (AED) task aims to identify and characterize anomalous events or patterns in temporal graphs and alert about them shortly after they start to occur. Since anomalies can manifest in many forms such as unexpected changes in the interaction patterns, shifts in network behavior, or the emergence of unusual group dynamics. Hence, the AED task's definition is closely related to the definition of an anomaly, in practice. Abstractly, we can assume the anomaly's definition is implicitly provided by the tagging of anomalies in a given dataset \cite{task_define_motivation}.

Mathematically, the AED task can be defined as follows. Let \(G\) be a temporal graph and let \(A = {a_1, a_2, \dots, a_n}\) represent the set of anomalies in \(G\) such that \(a_i := (U_i, T_i)\), where: \(U_i\) is a subset of nodes \(U_i \subset V\), representing the entities involved in the anomaly and \(T_i\) is a point in time that indicates the start of the anomaly emergence \(T_i \in T\). The AED task considered with finding a function \(M\) that accepts \(G\) and a subset \(A_{train} := (a_1, a_2, \dots, a_k)\) and predicts \(A_{test} := (a_{k+1}, \dots, a_n)\). 

For example, let us consider a temporal graph that represents a transportation network's dynamics, where nodes represent physical locations and edges represent the movement of vehicles between these locations, over time. An anomaly can be sudden and unexpected traffic congestion in a location or set of locations which could be caused by an accident or unplanned road closure. In this example, one can use historical records for such events and the data about the transportation network to try and predict the emergence of unexpected traffic congestion. 

\section{Experiment setup}
\label{sec:experiment}
In this section, we outline the experimental setup used for our benchmarking, including six main steps. 

\begin{figure}[!ht]
    \centering
    \includegraphics[width=0.99\textwidth]{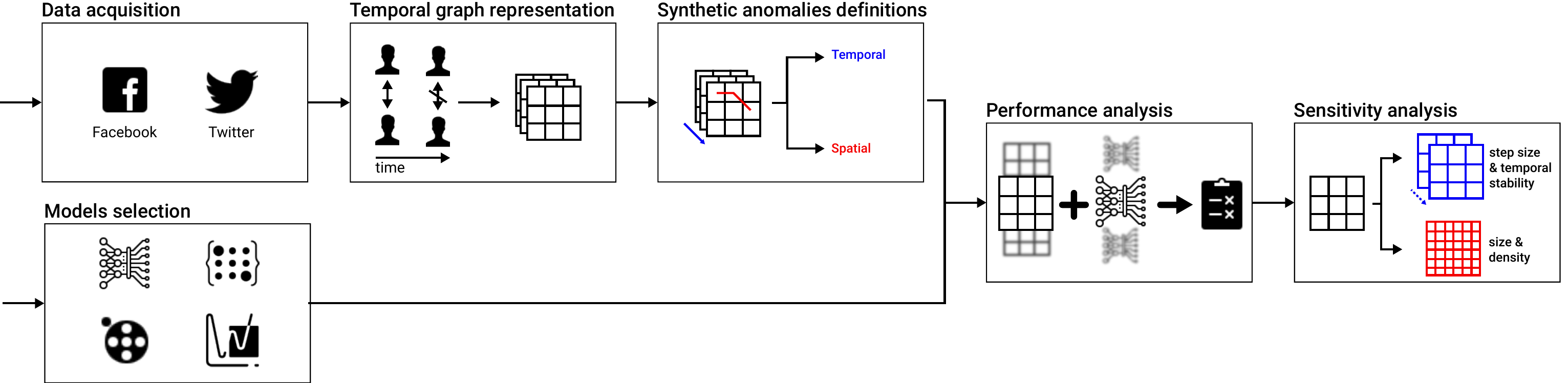}
    \caption{A schematic view of the experiments flow. First, we acquire data from a social media platform. Afterward, we represent the data as a temporal graph. Next, we define both temporal and spatial anomalies and test various models' performance in these settings. Finally, we conduct a sensitivity analysis on four properties of the temporal graph for each model.}
    \label{fig:enter-label}
\end{figure}

To conduct the benchmarking study, we carefully selected 12 data-driven models that encompass a wide range of computational approaches. Our aim was to ensure that these models represent the current state-of-the-art in the field, to the best of our knowledge. In the following sections, we provide a detailed description of each model, including its working principles and the rationale behind our selection.
\begin{itemize}
    \item Tree-based pipeline optimization tool (TPOT) \cite{tpot} - is an automated machine learning (AutoML) framework that optimizes a pipeline of preprocessing steps and machine learning models using genetic programming, based on the Scikit-learn library \cite{scikit}. 
    \item AutoKeras \cite{autokeras} - is an automated machine learning framework that uses neural architecture search to automatically select and optimize deep learning models based on the TensorFlow framework \cite{tensorflow}. 
    \item Time Series Anomaly Detection Using Generative Adversarial Networks (TADGAN) \cite{tadgan} - is a model that uses generative adversarial networks (GANs) to detect anomalies in time series data. We Include TADGAN in the analysis to explore the effectiveness of GAN framework for anomaly detection, which can capture both local and global patterns in the temporal graph data.
    \item Deep Isolation Forest (DIF) \cite{deep_isolation_forest} - is an extension of the Isolation Forest algorithm \cite{isolation_forest} that uses deep learning techniques to improve anomaly detection performance. 
    \item Long-short term memory (LSTM) neural network \cite{lstm} - is a type of recurrent neural network (RNN) that can model sequential data and capture long-term dependencies. It has the ability to learn temporal dependencies in the data without taking into consideration the graph-based nature of the data.
    \item Policy-based reinforcement learning for time series anomaly detection \cite{rl}. This model applies reinforcement learning techniques to train a policy network for anomaly detection in time series data. It is an adaptive approach that learns from a complex from a trial-and-error approach which potentially allows it the detection of complex and evolving anomalies.
    \item A XGboost for anomaly detection (XGBOD) \cite{xgboost_anomaly} - is an anomaly detection algorithm based on the XGBoost gradient boosting framework \cite{xgboost}. XGboost is widely considered one of the best machine learning models.
    \item A Python library for graph outlier detection (Pygod) \cite{pygod} - Pygod is a Python library specifically designed for detecting outliers in graph-structured data. 
    \item  Graph AutoEncoder with Random Forest (GAE\(+\)RF) \cite{graph_autoencoder,rf}. This model combines a graph autoencoder to obtain a meaningful representation of the data from the graph, operating as a feature engineering component that is used by an RF classifier. 
    \item Singular Value Decomposition with Random Forest (SVD\(+\)RF) \cite{svd,rf} - This model combines the singular value decomposition method which operates as an unsupervised feature engineering component followed by a random forest classifier. 
    \item Spatio-Temporal Graph Neural Networks (STGNN) \cite{stgnn} - is a model that integrates graph neural networks (GNNs) with spatial and temporal information for anomaly detection in spatio-temporal data. 
    \item Scalable Python Library for Time Series Data Mining (STUMPY) \cite{stumpy} - is a Python library that provides scalable algorithms for time series data mining, including motif discovery and time series approximation. 
\end{itemize}
In addition, we include a \textit{Random} model that just randomly decides if an anomaly occurs or not to be a naive baseline. 

We acquire data from the Twitter\footnote{\url{https://developer.twitter.com/en/docs/twitter-api}} and Facebook\footnote{\url{https://developers.facebook.com/docs/graph-api/}} social media websites using their official application programming interfaces (APIs). We picked these two social media websites as they provide access to the interaction data between their users over time. In addition to capturing user profiles, we also collected information about user interactions with posts (tweets) on both platforms. This included data on actions such as re-tweeting, commenting, and reacting (liking) to posts. For each interaction, we recorded the type of action, the timestamp, and the ID of the post owner. Overall, our dataset consisted of 44.8 thousand users from Twitter and 29.7 thousand users from Facebook, encompassing a total of 51.07 million and 65.93 million interactions, respectively. The data covered a duration of one month, specifically from the 22nd of August to the 22nd of September, 2020, and the 1st of February to the 1st of March, 2023, respectively.

In order to generate the temporal graphs representation of this data, one has to define the nodes and edges first. To this end, each account in the dataset represents a node, \(v \in V\) in the graph while an action (like, comment, share) that an account \(v \in V\) performance on a post of account \(u \in V\) at some time \(t \in \mathbb{N}\) represents an edge \(e := (v, u, t)\). Based on this definition, we obtain a direct temporal graph. For simplicity, we bin all actions to time durations of 15 minutes, in order to get a representation that agrees with a temporal sequence of graphs since the chosen models (see Section \ref{sec:models}) require such representation. 

Moreover, in order to obtain a population of temporal graphs from each dataset, we sampled 100 sub-graphs as follows. First, we picked at random a node of the graph, denoted by \(v_c\). Next, starting from \(v_c\), we computed Breadth-first search (BFS) \cite{bfs} while ignoring the time (\(t\)) component of the edges \(e \in E\) (and duplicate edges caused as a result) until \(|V| = 10000\) nodes are obtained. Once the nodes are obtained, we trimmed the temporal graph representing the entire dataset to include only these nodes. 

Since we do not have anomalies tagged on these temporal graphs, we had to generate them synthetically. Importantly, these synthetic tags have to be computed by information that is not fully available to the models; otherwise one would just examine the model's ability to reconstruct the rules used to generate the synthetic tags. As such, inspired by the works of \cite{anomlay_define_1,anomlay_define_2}, we define three anomaly rules. For all of them, let us consider a node \(v \in V\) at a time \(t \in \mathbb{N}\) to be anomaly if and only if:  \(N_t(v) > E_{t-z, t+z}[N(v)] + 2*S_{t-z, t+z}[N(v)] \) or \(\sum_{i = t-z}^{t+z} \frac{d^2N_i(v)}{di^2} > \sum_{i = t-z}^{t+z} \frac{1}{N_i(v)}\sum_{u \in C_i(v)}\frac{dN_i(u)}{di}\) or the largest eigenvalue of a matrix representing node's \(v\) number of interactions with the rest of nodes between \(t-z\) and \(t+z\) is larger than 1, where  \(C_t(v) := \{\forall u: (u, v, t) \in E\}\), \(N_t(v) := |C_t(v)|\), \(z \in \mathbb{N}\) is a window size, \(E_{a, b}(x)\) is the mean value of \(x\) such that \(t \in [a, b]\), and \(S_{a,b}(x)\) is the standard deviation value of \(x\) such that \(t \in [a, b]\).

Based on these anomalies, for each instance of a temporal graph, we computed the weighted \(F_1\) score \cite{f_one} using each one of the models. For all models, we used the first 80\% of temporal samples of each temporal graph instance to train the model while using the remaining 20\% for the evaluation. Importantly, the model's prediction is set to the next step in time, such that the window size is obtained for each model using the grid search method \cite{grid_search} ranging from \(1\) to \(2z\). 

Afterward, for each model, we conducted four sensitivity analysis tests, measuring the effect of changing one parameter of the task on each of the model's performances. Namely, the prediction lag, temporal concept drift, spatial size, and spatial density. Formally, we increase the prediction lag from 1 to \(z\) with steps of \(1\). For the temporal concept drift, for each step in time \(t\) with a probability \(p \in [0, 0.001, \dots, 0.01]\), all edges that are connected to node \(v\) are removed from the temporal graph. The spatial size sensitivity test was conducted by repeating the temporal graph instances construction but with \(9500 + 100i\) such that \(i \in [0, \dots, 10]\). Finally, the spatial was implemented by adding \(|E_0|t \cdot i \cdot 10^{-5} \) edges to the graph at time \(t\), where \(i \in [1, 10]\). 

\section{Results}
\label{sec:results}
\begin{figure}[!ht]
    \begin{subfigure}{.5\textwidth}
        \includegraphics[width=0.99\textwidth]{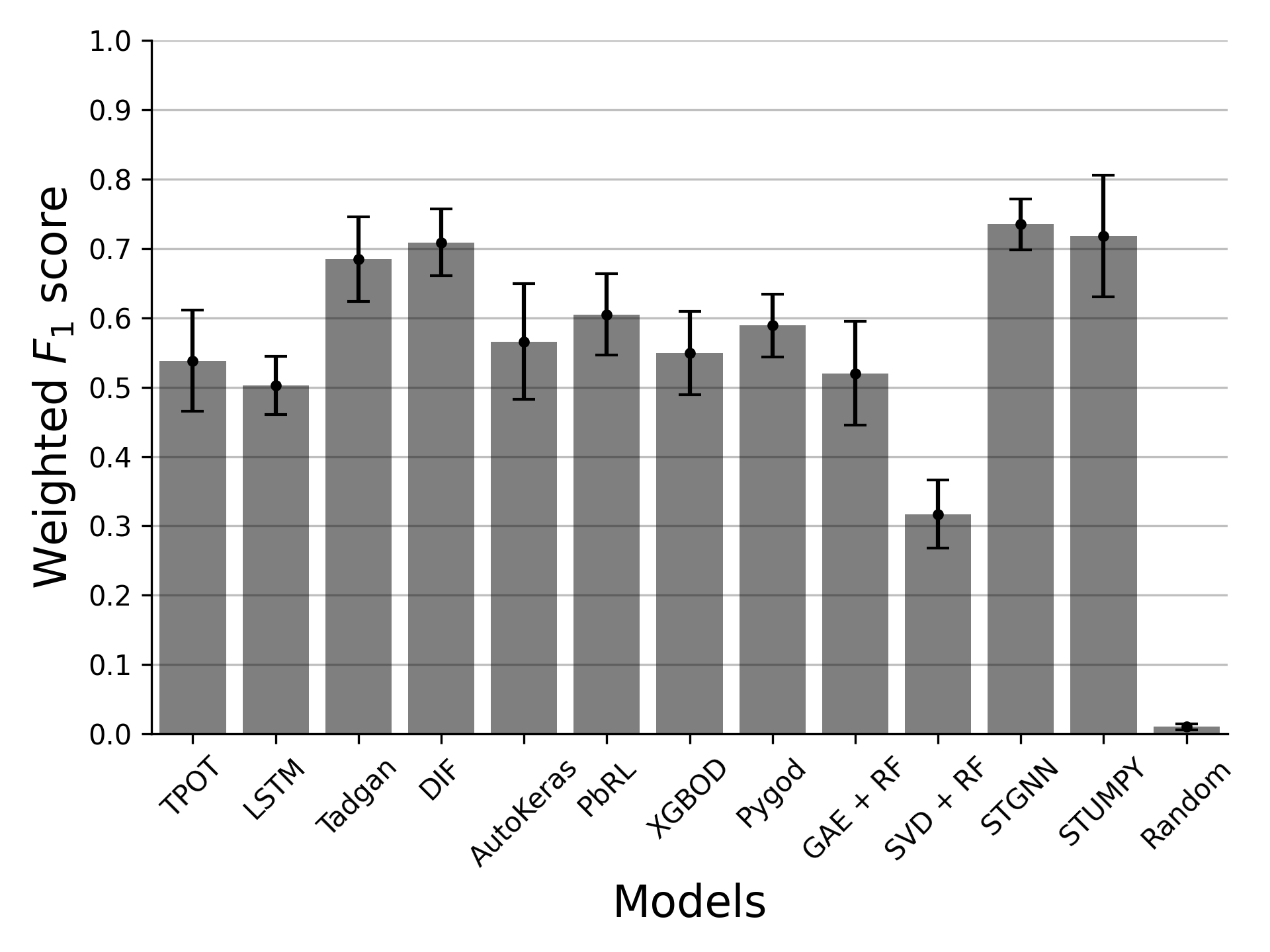}
        \caption{Twitter.}
        \label{fig:exp1}
    \end{subfigure}
    \begin{subfigure}{.5\textwidth}
        \includegraphics[width=0.99\textwidth]{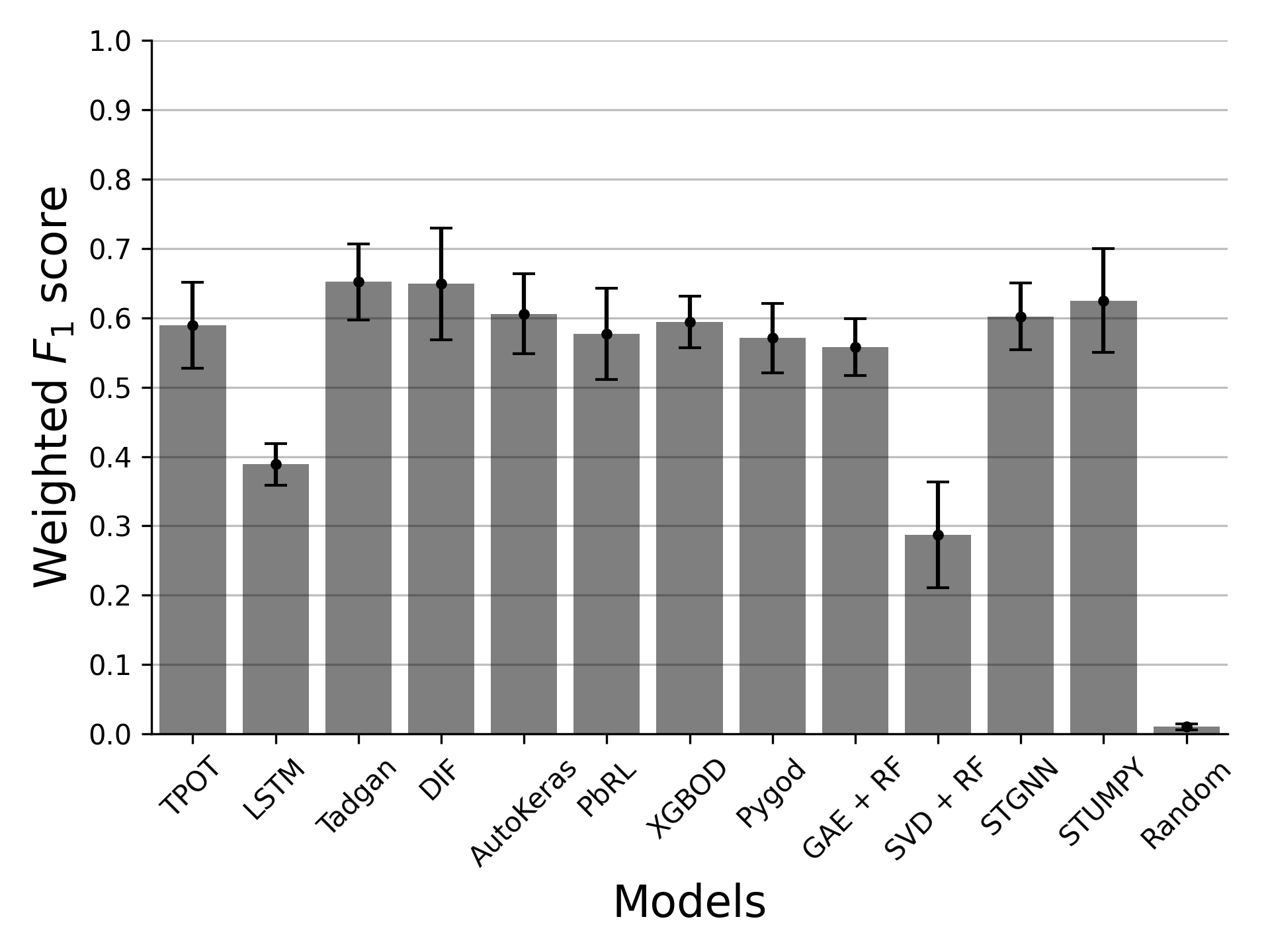}
        \caption{Facebook.}
        \label{fig:exp2}
    \end{subfigure}
    
    \caption{Comparison of different anomaly detection for the Twitter dataset.}
    \label{fig:exp}
\end{figure}

Fig. \ref{fig:exp} summarizes the main results obtained where Figs. \ref{fig:exp1} and \ref{fig:exp2} show the weighted \(F_1\) score of each model for the Twitter and Facebook datasets, respectively. The results are shown as the mean \(\pm\) standard deviation of \(n=100\) instances for each dataset. Upon examining the results, it becomes evident that the Facebook dataset consistently yielded lower performance, on average, compared to the Twitter dataset. This observation holds true when comparing each individual model's performance within the dataset, as well as when considering the collective performance of all the models. In addition, focusing on Fig. \ref{fig:exp1}, we can see that STGNN provides the best results with \(0.735 \pm 0.037\) followed by STUMPY with \(0.718 \pm 0.088\) and DIF with \(0.709 \pm 0.048\). All of the selected models in our benchmarking study are neural network-based approaches that have been specifically designed for anomaly detection. Unlike, Fig. \ref{fig:exp2} reveal that Tadgan obtained the best results with \(0.652 \pm 0.055\), followed by DIF with \(0.649 \pm 0.081\) and STUMPY with \(0.625 \pm 0.075\), showing somewhat consistency in the results. Similarly, the LSTM and SVD  with RF  models consistently performed worse compared to the other models. However, the performance order of the remaining models varied inconsistently between the two cases, indicating that the relative performance of these models is not consistently predictable or generalizable across different datasets or scenarios.

Furthermore, the sensitivity analysis results for each model have been summarized in Table \ref{table:sens}, which is divided into four sensitivity tests, and the values presented represent the average change in performance, as measured by the weighted \(F_1\) score, resulting from variations in the parameters investigated in each sensitivity test. 

\begin{table}[!ht]
\centering
\begin{tabular}
{p{0.2\textwidth}p{0.2\textwidth}p{0.2\textwidth}}
\hline \hline 
\textbf{Test}     & \textbf{Model} & \textbf{Value} \\ \hline \hline
\multirow{12}{*}{Prediction lag} & TPOT & \( -0.027 \) \\
& AutoKeras & \( -0.021 \) \\
& Tadgan  & \( \boldsymbol{-0.014} \) \\
& DIF  & \( -0.012 \) \\
& LSTM  & \( -0.030 \) \\
& Policy-based RL  & \( -0.017 \) \\
& XGBOD  & \( -0.018 \) \\
& Pygod  & \( -0.021 \) \\
& GAE + RF  & \( -0.025 \) \\
& SVD + RF  & \( -0.032 \) \\
& STGNN  & \( -0.015 \) \\
& STUMPY  & \( -0.015 \) \\ \hline
\multirow{12}{*}{Temopral concept drift} & TPOT & \( -0.052 \) \\
& AutoKeras & \( -0.032 \) \\
& Tadgan  & \( -0.038 \) \\
& DIF  & \( -0.041 \) \\
& LSTM  & \( -0.029 \) \\
& Policy-based RL  & \( -0.031 \) \\
& XGBOD  & \( -0.035 \) \\
& Pygod  & \( -0.040 \) \\
& GAE + RF  & \( \boldsymbol{-0.026} \) \\
& SVD + RF  & \( -0.028 \) \\
& STGNN  & \( -0.037 \) \\
& STUMPY  & \( -0.042 \) \\ \hline
\multirow{12}{*}{Spatial size} & TPOT & \( -0.007 \) \\
& AutoKeras & \( -0.008 \) \\
& Tadgan  & \( -0.011 \) \\
& DIF  & \( \boldsymbol{-0.006} \) \\
& LSTM  & \( -0.009 \) \\
& Policy-based RL  & \( -0.010 \) \\
& XGBOD  & \( -0.013 \) \\
& Pygod  & \( -0.012 \) \\
& GAE + RF  & \( -0.008 \) \\
& SVD + RF  & \( -0.008 \) \\
& STGNN  & \( -0.009 \) \\
& STUMPY  & \( -0.007 \) \\ \hline
\multirow{12}{*}{Spatial density} & TPOT & \( \boldsymbol{0.003} \) \\
& AutoKeras & \( -0.001 \) \\
& Tadgan  & \( -0.002 \) \\
& DIF  & \( -0.004 \) \\
& LSTM  & \( -0.002 \) \\
& Policy-based RL  & \( 0.002 \) \\
& XGBOD  & \( 0.005 \) \\
& Pygod  & \( 0.002 \) \\
& GAE + RF  & \( -0.001 \) \\
& SVD + RF  & \( -0.007 \) \\
& STGNN  & \( -0.002 \) \\
& STUMPY  & \( -0.002 \) \\ \hline \hline 
\end{tabular}
\caption{A sensitivity analysis of each model on four properties. The results are shown as an average change in the weighted \(F_1\) score. We marked in bold the best model for each sensitivity test.}
\label{table:sens}
\end{table}

\section{Discussion and Conclusion}
\label{sec:discussion}
In this study, we conducted a comprehensive benchmarking analysis to compare 12 data-driven methods for anomaly emergence detection in temporal graphs, with a specific focus on social media interactions. We evaluated the performance of these methods on two temporal graphs obtained from Twitter and Facebook, aiming to identify anomalies in pairwise and group interactions alike.

The comparison of various anomaly detection methods on both Twitter and Facebook datasets (Figs. \ref{fig:exp1} and \ref{fig:exp2}) reveals has yielded surprising results. Despite employing different computational approaches, several methods achieved statistically similar results while demonstrating inconsistency between the two datasets. This finding highlights the complex nature of anomaly detection in temporal graphs and the challenges associated with generalizing results across different platforms. For instance, we observed that the TPOT automatic machine-learning framework performed as the 9th-best model for the Twitter dataset, while ranking as the 7th-best for the Facebook dataset. This discrepancy emphasizes the need for tailored approaches and the consideration of dataset-specific characteristics when selecting the most effective anomaly detection method. Unsurprisingly, anomaly detection algorithms based on neural networks, such as STGNN and STUMPY  outperformed general-purpose models such as AutoKeras and LSTM-based neural networks. This outcome highlights the advantage of leveraging the inherent temporal dependencies and graph structures present in the data for improved anomaly detection performance. More generally, deep learning models seem to outperform other types of models. This can be explained by the ability of these models to capture more complex spatio-temporal connections in the data \cite{deep_good}. The inconsistency observed in the performance order of models across datasets further emphasizes the importance of dataset-specific exploration and evaluation. Different social media platforms exhibit unique characteristics in terms of user behaviors, network dynamics, and information propagation patterns. Indeed, the patterns of interactions differ between Twitter and Facebook significantly \cite{facebook_vs_twitter_1,facebook_vs_twitter_2}, leading to variations in the effectiveness of the methods. This outcome further supports the common no-free-lunch theorem as we were not able to find a single clear model that outperforms all others even on a small sample size of only two datasets \cite{free_lunch}. In the same manner, these results agree with a similar benchmarking analysis conducted for  unsupervised outlier node detection on static attributed graphs \cite{conlcusion}. More interestingly, Table \ref{table:sens} show that different models excel in different tests. Generally speaking, the models designed for anomaly detection are more sensitive to temporal concept drift and spatial density while for the prediction lag and spatial size, the generic purpose models were found to decrease in performance faster. This research contributes to a better understanding of the complexities and challenges associated with anomaly detection in large and dynamic systems represented as temporal graphs. Future work should continue to explore novel techniques and methodologies that can effectively address these challenges and provide more robust anomaly detection solutions for diverse real-world applications.

This study is not without limitations. First, the evaluation was conducted on a limited number of datasets, which may not fully capture the diversity and complexity of social media interactions. Furthermore, the anomalies used in this study are synthetic due to the time and resource burden of tagging such events in real data. As such, our results might change slightly given realistic anomaly tagging. Commonly, data-driven models in general, and anomaly detection models, in particular, benefit from the introduction of domain knowledge \cite{teddy_ae_ki,ae_info_1,ae_info_2,scimed,ae_info_intro}. As such, it is of great interest how the proposed results would alter if domain knowledge is integrated into the examined models. These limitations propose a fertile soil for future studies of temporal graph anomaly emergence detection. 

\section*{Declarations}
\subsection*{Funding}
This research did not receive any specific grant from funding agencies in the public, commercial, or not-for-profit sectors.

\subsection*{Conflicts of interest/Competing interests}
None.

\subsection*{Data availability}
The data used as part of this study is available upon reasonable request from the authors.

\subsection*{Acknowledgement}
The author wishes to thank Tom Hope for inspiring this research and implicitly suggesting several of the models used as part of this study. 

\subsection*{Author Contribution}
Teddy Lazebnik: Conceptualization, Data Curation, Methodology, Software, Validation, Formal analysis, Investigation, Writing - Original Draft, Writing - Review \& Editing. \\
Or Iny: Conceptualization, Data Curation, Validation, Resources, Writing - Review \& Editing.
 
\bibliography{biblio}

\begin{thebibliography}{10}

\bibitem{intro_social}
G.~Robins and P.~Pattison.
\newblock Random graph models for temporal processes in social networks.
\newblock {\em The Journal of Mathematical Sociology}, 25(1):5--41, 2001.

\bibitem{intro_bio}
M.~Zheng, S.~Domanskyi, C.~Piermarocchi, and G.~I. Mais.
\newblock Visibility graph based temporal community detection with applications
  in biological time series.
\newblock {\em Scientific Reports}, 11:5623, 2021.

\bibitem{intro_cars}
G.~Del~Mondo, P.~Peng, J.~Gensel, C.~Claramunt, and F.~Lu.
\newblock Leveraging spatio-temporal graphs and knowledge graphs: Perspectives
  in the field of maritime transportation.
\newblock {\em ISPRS International Journal of Geo-Information}, 10(8), 2021.

\bibitem{tg_tag_1}
L.~Zhao, Y.~Song, C.~Zhang, Y.~Liu, P.~Wang, T.~Lin, M.~Deng, and H.~Li.
\newblock T-gcn: A temporal graph convolutional network for traffic prediction.
\newblock {\em IEEE Transactions on Intelligent Transportation Systems},
  21(9):3848--3858, 2020.

\bibitem{tg_tag_2}
X.~Wang, Y.~Ma, Y.~Wang, W.~Jin, X.~Wang, J.~Tang, C.~Jia, and J.~Yu.
\newblock Traffic flow prediction via spatial temporal graph neural network.
\newblock In {\em Proceedings of The Web Conference 2020}, page 1082–1092.
  Association for Computing Machinery, 2020.

\bibitem{tg_tag_3}
G.~Xiao, R.~Wang, C.~Zhang, and A.~Ni.
\newblock Demand prediction for a public bike sharing program based on
  spatio-temporal graph convolutional networks.
\newblock {\em Multimedia Tools and Applications}, 80, 2021.

\bibitem{tg_tag_4}
C.~Zhang, J.~J.~Q. Yu, and Y.~Liu.
\newblock Spatial-temporal graph attention networks: A deep learning approach
  for traffic forecasting.
\newblock {\em IEEE Access}, 7:166246--166256, 2019.

\bibitem{tg_math_1}
S.~Huang, J.~Cheng, and H.~Wu.
\newblock Temporal graph traversals: Definitions, algorithms, and applications.
\newblock {\em arXiv}, 2014.

\bibitem{tg_anomaly_1}
L.~Cai, Z.~Chen, C.~Luo, J.~Gui, J.~Ni, D.~Li, and H.~Chen.
\newblock Structural temporal graph neural networks for anomaly detection in
  dynamic graphs.
\newblock In {\em Proceedings of the 30th ACM International Conference on
  Information \& Knowledge Management}, page 3747–3756, 2021.

\bibitem{tg_anomaly_2}
S.~Rayana and L.~Akoglu.
\newblock {\em Less is More: Building Selective Anomaly Ensembles with
  Application to Event Detection in Temporal Graphs}, page 622.
\newblock Proceedings of the 2015 SIAM International Conference on Data Mining,
  2015.

\bibitem{tg_ano_exp_1}
D.~Cao, Y.~Wang, J.~Duan, C.~Zhang, X.~Zhu, C.~Huang, Y.~Tong, B.~Xu, J.~Bai,
  J.~Tong, and Q.~Zhang.
\newblock Spectral temporal graph neural network for multivariate time-series
  forecasting.
\newblock In {\em Advances in Neural Information Processing Systems},
  volume~33, pages 17766--17778, 2020.

\bibitem{tg_ano_exp_2}
W.~Chung and V.~S. Lai.
\newblock A temporal graph framework for intelligence extraction in social
  media networks.
\newblock {\em Information \& Management}, 60(4):103773, 2023.

\bibitem{tg_ano_exp_3}
D.~Fu, L.~Fang, R.~Maciejewski, V.~I. Torvik, and J.~He.
\newblock Meta-learned metrics over multi-evolution temporal graphs.
\newblock In {\em Proceedings of the 28th ACM SIGKDD Conference on Knowledge
  Discovery and Data Mining}, page 367–377, 2022.

\bibitem{ano_task_1}
H.~Du, S.~Wang, and H.~Huo.
\newblock Xfinder: Detecting unknown anomalies in distributed machine learning
  scenario.
\newblock {\em Front. Comput. Sci.}, 3, 2021.

\bibitem{ano_task_2}
D.~Liu, Y.~Zhao, H.~Xu, Y.~Sun, D.~Pei, J.~Luo, X.~Jing, and M.~Feng.
\newblock Opprentice: Towards practical and automatic anomaly detection through
  machine learning.
\newblock In {\em Proceedings of the 2015 Internet Measurement Conference},
  page 211–224, 2015.

\bibitem{ano_task_3}
C.~Ding, S.~Sun, and J.~Zhao.
\newblock Mst-gat: A multimodal spatial–temporal graph attention network for
  time series anomaly detection.
\newblock {\em Information Fusion}, 89:527--536, 2023.

\bibitem{ano_task_4}
X.~Zeng, Y.~Jiang, W.~Ding, H.~Li, Y.~Hao, and Z.~Qiu.
\newblock A hierarchical spatio-temporal graph convolutional neural network for
  anomaly detection in videos.
\newblock {\em IEEE Transactions on Circuits and Systems for Video Technology},
  33(1):200--212, 2023.

\bibitem{ano_task_5}
L.~Cai, Z.~Chen, C.~Luo, J.~Gui, J.~Ni, D.~Li, and H.~Chen.
\newblock Structural temporal graph neural networks for anomaly detection in
  dynamic graphs.
\newblock In {\em Proceedings of the 30th ACM International Conference on
  Information \& Knowledge Management}, page 3747–3756, 2021.

\bibitem{ano_task_6}
S.~Pandhre, H.~Mittal, M.~Gupta, and V.~N. Balasubramanian.
\newblock Stwalk: Learning trajectory representations in temporal graphs.
\newblock In {\em Proceedings of the ACM India Joint International Conference
  on Data Science and Management of Data}, page 210–219, 2018.

\bibitem{rw_intro_1}
L.~F.~A. Brito, B.~A.~N. Travencolo, and M.~K. Alertini.
\newblock A review of in-memory space-efficient data structures for temporal
  graphs.
\newblock {\em arXiv}, 2022.

\bibitem{rw_intro_2}
P.~Holme and J.~Saramaki.
\newblock Temporal networks.
\newblock {\em Physics Reports}, 519(3):97--125, 2012.

\bibitem{rw_intro_3}
T.~Zhang, Y.~Gao, L.~Qiu, L.~Chen, Q.~Linghu, and S.~Pu.
\newblock Distributed time-respecting flow graph pattern matching on temporal
  graphs.
\newblock {\em World Wide Web}, 23:609--630, 2020.

\bibitem{tg_intro_book}
O.~Michail.
\newblock An introduction to temporal graphs: An algorithmic perspective.
\newblock {\em arXiv}, 2015.

\bibitem{rw_genes}
M.~J. McNeil, L.~Zhang, and P.~Bogdanov.
\newblock Temporal graph signal decomposition.
\newblock In {\em Proceedings of the 27th ACM SIGKDD Conference on Knowledge
  Discovery \& Data Mining}, page 1191–1201, 2021.

\bibitem{rw_gene_exp}
M.~Bansal and D.~di~Bernardo.
\newblock Inference of gene networks from temporal gene expression profiles.
\newblock {\em IET Systems Biology}, 1:306--312(6), 2007.

\bibitem{rw_traffic_example}
Q.~Zhang, J.~Chang, G.~Meng, S.~Xiang, and C.~Pan.
\newblock Spatio-temporal graph structure learning for traffic forecasting.
\newblock {\em Proceedings of the AAAI Conference on Artificial Intelligence},
  34(01):1177--1185, 2020.

\bibitem{rw_social_general_1}
J.~Byun, S.~Woo, and D.~Kim.
\newblock Chronograph: Enabling temporal graph traversals for efficient
  information diffusion analysis over time.
\newblock {\em IEEE Transactions on Knowledge and Data Engineering},
  32(3):424--437, 2020.

\bibitem{rw_social_general_2}
S.~K. Maity, T.~V. Manoj, and A.~Mukherjee.
\newblock Opinion formation in time-varying social networks: The case of the
  naming game.
\newblock {\em Phys. Rev. E}, 86:036110, 2012.

\bibitem{rw_social_example}
J.~Plepi, F.~Sakketou, H-J. Geiss, and L.~Flek.
\newblock Temporal graph analysis of misinformation spreaders in social media.
\newblock In {\em Proceedings of TextGraphs-16: Graph-based Methods for Natural
  Language Processing}, pages 89--104, 2022.

\bibitem{task_define_motivation}
A.~Bl\'{a}zquez-Garc\'{\i}a, A.~Conde, U.~Mori, and J.~A. Lozano.
\newblock A review on outlier/anomaly detection in time series data.
\newblock {\em ACM Comput. Surv.}, 54(3):56, 2021.

\bibitem{tpot}
R.~S. Olson and J.~H. Moore.
\newblock Tpot: A tree-based pipeline optimization tool for automating machine
  learning.
\newblock In {\em Workshop on automatic machine learning}, pages 66--74. PMLR,
  2016.

\bibitem{scikit}
F.~Pedregosa, G.~Varoquaux, A.~Gramfort, V.~Michel, B.~Thirion, O.~Grisel,
  M.~Blondel, P.~Prettenhofer, R.~Weiss, V.~Dubourg, J.~Vanderplas, A.~Passos,
  D.~Cournapeau, M.~Brucher, M.~Perrot, and E.~Duchesnay.
\newblock Scikit-learn: Machine learning in {P}ython.
\newblock {\em Journal of Machine Learning Research}, 12:2825--2830, 2011.

\bibitem{autokeras}
H.~Jin, F.~Chollet, Q.~Song, and X.~Hu.
\newblock Autokeras: An automl library for deep learning.
\newblock {\em Journal of Machine Learning Research}, 24(6):1--6, 2023.

\bibitem{tensorflow}
M.~Abadi, P.~Barham, J.~Chen, Z.~Chen, A.~Davis, J.~Dean, M.~Devin,
  S.~Ghemawat, G.~Irving, and M.~Isard.
\newblock Tensorflow: A system for large-scale machine learning.
\newblock In {\em 12th $\{$USENIX$\}$ Symposium on Operating Systems Design and
  Implementation ($\{$OSDI$\}$ 16)}, pages 265--283, 2016.

\bibitem{tadgan}
A.~Geiger, D.~Liu, S.~Alnegheimish, A.~Cuesta-Infante, and K.~Veeramachaneni.
\newblock Tadgan: Time series anomaly detection using generative adversarial
  networks.
\newblock {\em arXiv}, 2020.

\bibitem{deep_isolation_forest}
H.~Xu, G.~Pang, Y.~Wang, and Y.~Wang.
\newblock Deep isolation forest for anomaly detection.
\newblock {\em arXiv}, 2023.

\bibitem{isolation_forest}
F.~T. Liu, K.~M. Ting, and Z-H. Zhou.
\newblock Isolation forest.
\newblock In {\em Data Mining}, pages 265--283. ICDM’08, 2008.

\bibitem{lstm}
I.~Sutskever, O.~Vinyals, and Q.~V. Le.
\newblock Sequence to sequence learning with neural networks.
\newblock {\em Advances in Neural Information Processing Systems},
  27:3104--3112, 2014.

\bibitem{rl}
M.~Yu and S.~Sun.
\newblock Policy-based reinforcement learning for time series anomaly
  detection.
\newblock {\em Engineering Applications of Artificial Intelligence}, 95:103919,
  2020.

\bibitem{xgboost_anomaly}
Y.~Zhao and M.~K. Hryniewicki.
\newblock Xgbod: Improving supervised outlier detection with unsupervised
  representation learning.
\newblock {\em arXiv}, 2019.

\bibitem{xgboost}
T.~Chen and C.~Guestrin.
\newblock {XGBoost}: A scalable tree boosting system.
\newblock In {\em Proceedings of the 22nd ACM SIGKDD International Conference
  on Knowledge Discovery and Data Mining}, KDD '16, pages 785--794. ACM, 2016.

\bibitem{pygod}
K.~Liu, Y.~Dou, Y.~Zhao, X.~Ding, X.~Hu, R.~Zhang, K.~Ding, C.~Chen, H.~Peng,
  K.~Shu, G.~H. Chen, Z.~Jia, and P.~S. Yu.
\newblock Pygod: A python library for graph outlier detection.
\newblock {\em arXiv}, 2022.

\bibitem{graph_autoencoder}
T.~N. Kipf and M.~Welling.
\newblock Variational graph auto-encoders.
\newblock {\em NIPS Workshop on Bayesian Deep Learning}, 2016.

\bibitem{rf}
T.~K. Ho.
\newblock Random decision forests.
\newblock In {\em Proceedings of 3rd international conference on document
  analysis and recognition}, volume~1, pages 278--282. IEEE, 1995.

\bibitem{svd}
V.~Klema and A.~Laub.
\newblock The singular value decomposition: Its computation and some
  applications.
\newblock {\em IEEE Transactions on Automatic Control}, 25(2):164--176, 1980.

\bibitem{stgnn}
J.~Chen, Y.~Wang, R.~Wu, and M.~Campbell.
\newblock Spatial-temporal graph neural network for interaction-aware vehicle
  trajectory prediction.
\newblock In {\em 2021 IEEE 17th International Conference on Automation Science
  and Engineering (CASE)}, pages 2119--2125, 2021.

\bibitem{stumpy}
S.~M. Law.
\newblock {STUMPY: A Powerful and Scalable Python Library for Time Series Data
  Mining}.
\newblock {\em {The Journal of Open Source Software}}, 4(39):1504, 2019.

\bibitem{bfs}
D.~C. Kozen.
\newblock {\em Depth-First and Breadth-First Search}, pages 19--24.
\newblock Springer New York, 1992.

\bibitem{anomlay_define_1}
R.~Yu, H.~Qiu, Z.~Wen, C.~Lin, and Y.~Liu.
\newblock A survey on social media anomaly detection.
\newblock {\em SIGKDD Explor. Newsl.}, 18(1):1–14, 2016.

\bibitem{anomlay_define_2}
R.~Yu, X.~He, and Y.~Liu.
\newblock Glad: Group anomaly detection in social media analysis.
\newblock {\em ACM Trans. Knowl. Discov. Data}, 10(2), 2015.

\bibitem{f_one}
C.~Goutte and E.~Gaussier.
\newblock A probabilistic interpretation of precision, recall and f-score, with
  implication for evaluation.
\newblock In D.~E. Losada and J.~M. Fernandez-Luna, editors, {\em Advances in
  Information Retrieval}, pages 345--359. Springer Berlin Heidelberg, 2005.

\bibitem{grid_search}
R.~Liu, E.~Liu, J.~Yang, M.~Li, and F.~Wang.
\newblock Optimizing the hyper-parameters for svm by combining evolution
  strategies with a grid search.
\newblock {\em Intelligent Control and Automation}, 344, 2006.

\bibitem{deep_good}
C.~Janiesch, P.~Zschech, and K.~Heinrich.
\newblock Machine learning and deep learning.
\newblock {\em Electron Markets}, 31:685--695, 2021.

\bibitem{facebook_vs_twitter_1}
K.~Jaidka, S.~Guntuku, and L.~Ungar.
\newblock Facebook versus twitter: Differences in self-disclosure and trait
  prediction.
\newblock {\em Proceedings of the International AAAI Conference on Web and
  Social Media}, 12(1), 2018.

\bibitem{facebook_vs_twitter_2}
N.~Petrocchi, A.~Asnaani, A.~P. Martinez, A.~Nadkarni, and S.~G. Hofmann.
\newblock Differences between people who use only facebook and those who use
  facebook plus twitter.
\newblock {\em International Journal of Human–Computer Interaction},
  31(2):157--165, 2015.

\bibitem{free_lunch}
D.~H. Wolpert and W.~G. Macready.
\newblock No free lunch theorems for optimization.
\newblock {\em IEEE Transactions on Evolutionary Computation}, 67, 1997.

\bibitem{conlcusion}
K.~Liu, Y.~Dou, Y.~Zhao, X.~Ding, X.~Hu, R.~Zhang, K.~Ding, C.~Chen, H.~Peng,
  K.~Shu, L.~Sun, J.~Li, G.~H. Chen, Z.~Jia, and P.~S. Yu.
\newblock Bond: Benchmarking unsupervised outlier node detection on static
  attributed graphs.
\newblock {\em Advances in Neural Information Processing Systems},
  35:27021--27035, 2022.

\bibitem{teddy_ae_ki}
T.~Lazebnik and L.~Simon-Keren.
\newblock Knowledge-integrated autoencoder model.
\newblock {\em arXiv}, 2023.

\bibitem{ae_info_1}
T.~Ma and A.~Zhang.
\newblock Integrate multi-omics data with biological interaction networks using
  multi-view factorization autoencoder (mae).
\newblock {\em BMC Genomics}, 20:944, 2019.

\bibitem{ae_info_2}
W.~Ding, H.~Lin, B.~Li, K.~J. Eun, and D.~Zhao.
\newblock Semantically adversarial driving scenario generation with explicit
  knowledge integration.
\newblock {\em arXiv}, 2022.

\bibitem{scimed}
L.~S. Keren, A.~Liberzon, and T.~Lazebnik.
\newblock A computational framework for physics-informed symbolic regression
  with straightforward integration of domain knowledge.
\newblock {\em Scientific Reports}, 13(1):1249, 2023.

\bibitem{ae_info_intro}
Y.~Deng, A.~Sander, L.~Faulstich, and K.~Denecke.
\newblock Towards automatic encoding of medical procedures using convolutional
  neural networks and autoencoders.
\newblock {\em Artificial Intelligence in Medicine}, 93:29--42, 2019.

\end{thebibliography}
\bibliographystyle{unsrt}

\end{document}